\documentclass[iop,twocolappendix,revtex4]{emulateapj}
\usepackage{graphicx,natbib,enumitem}
\usepackage{amsmath,amssymb,mathrsfs,bm}
\usepackage{color,txfonts}

\def\fig{Figure~}

\def\equ{Equation~}

\def\part{Section~}

\def\planck{\textit{Planck}{}}

\def\Asz{\mathrm{A_{\mathrm{SZ}}}}
\def\Acmb{\mathrm{A_{\mathrm{CMB}}}}
\def\Adust{\mathrm{A_{\mathrm{dust}}}}

\def\fsz{{f_{\mathrm{SZ}}}}

\def\fdust{{f_{\mathrm{dust}}}}

\def\cur{\mathscr{Cur}}
\def\wt{\mathscr{Wt}}
\def\rec{\mathcal{R}}

\def\reg{\mathcal{S}}
\def\thetavec{\boldsymbol{\theta}}

\def\bthreespline{$\mathrm{{B}_{3}}$-spline{~}}

\newcommand{\vect}[1]{\boldsymbol{{#1}}}

\newcommand{\myemail}{herve.bourdin@roma2.infn.it}

\shorttitle{SZ spectral imaging with \textit{Planck}}
\shortauthors{Bourdin, Mazzotta, \& Rasia}

\begin{document}

\title{Spectral imaging of galaxy clusters with \textit{Planck}}
\author{H. Bourdin\altaffilmark{1}, P. Mazzotta \altaffilmark{1,2}, E. Rasia\altaffilmark{3,4}}
\journalinfo{The Astrophysical Journal, 815:92 (11pp), 2015 December 20}
\submitted{Received 2015 August 3; accepted 2015 October 30; published 2015 December 14}

\affil{{$^1$}{Dipartimento di Fisica, Universit\`a degli Studi di Roma `Tor
  Vergata', via della Ricerca Scientifica, 1, I-00133 Roma, Italy; \myemail}}
\affil{{$^2$}{Harvard Smithsonian Centre for Astrophysics, 60 Garden Street, 
Cambridge, MA 02138, USA}}
\affil{{$^3$}{INAF-Osservatorio Astronomico of Trieste, via Tiepolo 11, I-34121 Trieste, Italy}}
\affil{{$^4$}{Department of Physics, University of Michigan, Ann Arbor, MI 48109, USA}}

\begin{abstract}

The Sunyaev-Zeldovich (SZ) effect is a promising tool for detecting the presence of hot gas out to the galaxy cluster peripheries. We developed a spectral imaging algorithm dedicated to the SZ observations of nearby galaxy clusters with \planck, with the aim of revealing gas density anisotropies related to the filamentary accretion of materials, or pressure discontinuities induced by the propagation of shock fronts. To optimize an unavoidable trade-off between angular resolution and precision of the SZ flux measurements, the algorithm performs a multiscale analysis of the SZ maps as well as of other extended components, such as the cosmic microwave background (CMB) anisotropies and the Galactic thermal dust. The demixing of the SZ signal is tackled through kernel weighted likelihood maximizations. The CMB anisotropies are further analyzed through a wavelet analysis, while the Galactic foregrounds and SZ maps are analyzed via a curvelet analysis that best preserves their anisotropic details. The algorithm performance has been tested against mock observations of galaxy clusters obtained by simulating the \planck~ High Frequency Instrument and by pointing a few characteristic positions in the sky. These tests suggest that \planck~  should easily allow us to detect filaments in the cluster peripheries and detect large-scale shocks in colliding galaxy clusters that feature favorable geometry.

\end{abstract}


\section{Introduction}

Most of the baryonic content of galaxy clusters is in the form of a hot ionized gas that is in pressure equilibrium with gravity. This intracluster medium (ICM) is detectable in X-rays via its bremsstrahlung emission, as well as in the millimetric band via the inverse Compton diffusion of the cosmic microwave background (CMB) radiation, or the so-called Sunyaev-Zeldovich (hereafter SZ) effect. The excellent agreement between X-ray and SZ measurements of the central region of galaxy clusters, $r \leq r_{500}$ \footnote{$r_{\Delta}$ is the radius of a ball whose density is $\Delta$ times the critical density of the universe}, suggests that both observables accurately probe the integrated ICM thermal pressure \citep[see, e.g.][]{Planck_early_XI_SZscaling}. Due to the quadratic dependence of the X-ray emission measure on the ICM density, X-ray observations mostly enlighten the physics of the innermost cluster regions ($r < r_{2500}$), while X-ray spectroscopy poorly constrains the outer regions. By contrast, the SZ Compton parameter is proportional to the integrated ICM pressure along the line of sight. Therefore, the SZ signal extends further out and enables us to explore the complex baryonic physics at play in the cluster outskirts.

\planck~ is the third space satellite that has mapped the CMB over the full-sky. Its high sensitivity and unprecedented angular resolution allowed the detection of more than 1000 galaxy clusters classified in the Second \planck~ Catalogue SZ sources \citep{Planck_2015_szcatalogue}. Releases of \planck~ data proved the strength of the SZ effect in detecting the ICM out to the cluster peripheries.  In particular, radially averaged measurements of the Compton parameter of 60 nearby massive clusters showed that their averaged pressure profile is similar to X-ray derived profiles below $r_{500}$, but slightly exceeds the theoretical predictions from cosmological simulations of cluster formation at $r \ge r_{500}$ \citep{Planck_2013_V_pressure}. A subsequent combination of  \planck~ and ROSAT data toward 18 of these clusters revealed the different natures of the hot gas entropy profiles of the cool core and non-cool core clusters at $r_{200}$ \citep{Eckert_13a}, and also managed to discriminate between averaged gas fractions in the outer regions of the two cluster classes \citep{Eckert_13b}. Complementary to radial profiles, maps of the Compton parameter provide us with a precious information that helps with the interpretation of these results.  Indeed, SZ maps allow us to identify merger shocks that locally raise the ICM entropy and accretion filaments that affect the ICM hydrostatic equilibrium via the anisotropic injection of turbulence. From an observational point of view, fine two-dimensional (2D) information helps with the identification of some sources of bias of the radially averaged SZ and X-ray flux measurements generated by structures projected along the line of sight. Furthermore, the detection of accretion filaments also identifies ICM regions that are likely inhomogeneous \citep[e.g.][]{Vazza_13}, yielding nonlinear biases when assuming spherical symmetry in the inversion of the gas density from X-ray surface brightness profiles \citep{Nagai_11}. The first analyses of Planck data have already allowed us to detect highly significant anisotropies in the cluster atmospheres, in particular two shock fronts in the Coma cluster \citep{Planck_intermediate_X_coma} and a Mpc-scale filament connecting both components of the cluster pair A399-A401 \citep{Planck_intermediate_VIII_filaments}. These results encourage us to develop an image restoration algorithm that could reveal SZ anisotropies down to a lower signal-to-noise ratio.

The SZ signal from galaxy clusters is a mixture of up to several extended emissions of Galactic and extragalactic origin that can be demixed through local likelihood maximizations. By taking advantage of kernel-weighted $\chi^2$ minimisations, we propose to optimize the bias-variance trade-off of these estimates via a multi-scale analysis. Including a wavelet analysis of the CMB map and a curvelet analysis of the foreground maps, the proposed algorithm has been adapted to match the spectral responses and beams of the Planck High Frequency Instrument (HFI), and includes, in particular, an iterative deconvolution of the frequency maps. After having detailed our algorithm in Section 2, we present mock HFI observations of galaxy clusters and discuss the algorithm performance in mapping the SZ signal from these simulations in Section 3. In the following analysis, intracluster distances are computed as angular diameter distances, assuming a $\Lambda$-CDM cosmology with ${H}_\mathrm{0} =70~\mathrm{km}~\mathrm{s}^{-1}\mathrm{Mpc}^{-1}$, $\Omega_{\mathrm{M}} = 0.3$, $\Omega_{\Lambda} = 0.7$. 

\begin{figure*}[ht]
   \begin{center}
  \resizebox{\hsize}{!}{\includegraphics{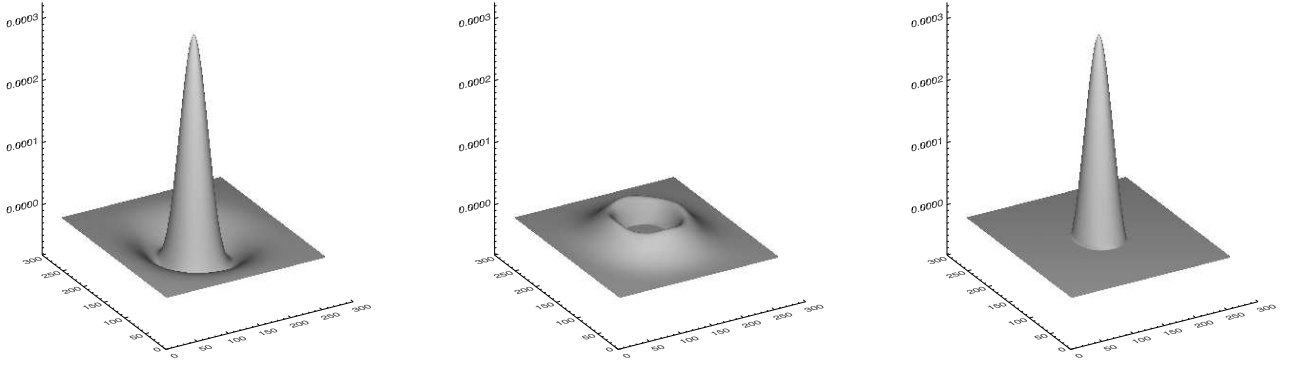}}
  \caption{Left: Isotropic \bthreespline wavelet function $\psi$. Middle and right: negative and positive parts of the wavelet function, $\psi_{-}$ and $\psi_{+}$ \label{fig:bspline_kernel}}
   \end{center}
\end{figure*}

\section{Spectral imaging of the thermal SZ signal\label{sect:spectral-imaging}}

\subsection{Toward spectral imaging}

Across the radio electromagnetic spectrum, the SZ signal is a mixture of CMB radiation, and several extended emissions of Galactic (free-free, synchrotron and thermal dust continua, CO lines) and extragalactic origin (Cosmic Infrared Background). When looking at all-sky gigahertzian frequency maps $\vect{f}$ that are contaminated by their instrumental noise $\vect{N}$, separating these components is an inverse problem that is traditionally stated as:
\begin{equation}
\vect{f} = \vect{As} + \vect{N},
\label{equ:component_separation}
\end{equation}
where we look for a mixing matrix $\vect{A}$ between several unknown emitting sources $\vect{s}$. Originally motivated by the denoising of all-sky CMB maps, a number of component separation algorithms have been proposed for solving Equ.(\ref{equ:component_separation}): 

\begin{enumerate}[leftmargin=.75cm,noitemsep]
\item Parametric methods where the spectral energy distributions (SEDs) of all components are used as a priori knowledge to maximize their spatial distribution entropy or likelihood \citep{Hobson_98,Stolyarov_05,Eriksen_08,Khatri_15};
\item Internal linear combination methods that map a specific component as a linear combination of the observed maps with the constraint of minimizing its variance \citep{Eriksen_04,Remazeilles_11,Hurier_13}; and
\item Blind or semi-blind source separation methods that rely on spatial source independence or sparsity priors \citep{Delabrouille_03,Cardoso_08,Bobin_13}. 
\end{enumerate}

Unlike the image of the last scattering surface, some of the foreground sources are localized objects that may not be uniformly modeled by Gaussian random fields. Therefore, one of the challenges of the component separation algorithms is to avoid any unreal amplification of the very low or null amplitude elements of the source vector, $\vect{s}$. In this context, parametric methods have shown their robustness for mapping the brightest extended foregrounds \citep[see, e.g.][]{Planck_2013_XI_tdustmodel}, while sparsity-regularized reconstruction approaches have proven their ability to spatially separate the CMB anisotropies from fainter and more localized objects \citep{Bobin_14}. 

Among sparse representations, wavelet transforms have long been successfully used to denoise X-ray images of galaxy clusters \citep{Slezak_94,Vikhlinin_97,Starck_98}, and thus have naturally been proposed for SZ imaging \citep{Pierpaoli_05,Pires_06}. The isotropy of 2D wavelet functions, however, is unsuitable for detecting and preserving the filamentary structures that are likely to populate the  cluster outskirts as well as the the Galactic dust. A solution can be represented by curvelets that are especially designed to perform a sparse representation of linear and curved edges in the images \citep{Candes_00}. To combine the local robustness of parametric methods to the imaging capabilities offered by sparse representations, we propose demixing the SZ signal with respect to the CMB and thermal dust via a set of kernel-weighted likelihood maximisations that are directly related to the wavelet transform of the CMB map and to the curvelet transforms of the foreground maps.

\subsection{Component estimate}

The \planck~HFI is composed of bolometer arrays whose frequency channels cover the most prominent part of the thermal SZ (tSZ) energy spectrum. In the HFI frequency range (100--857 GHz), the tSZ component is locally contaminated by radio and infrared point sources, and is mostly mixed up with the CMB temperature anisotropies and with the Galactic extended emissions from the thermal dust and CO lines \citep{Planck_2015_X_foregrounds}. Moreover, most of the time the latter component is negligible at high Galactic latitude, as it is a tracer of Galactic molecular clouds. For these reasons, we reduce the specific intensity expected for each pixel ($k$,$l$) of the HFI frequency maps to the sum of three SEDs associated with the tSZ, thermal dust, and CMB components respectively,

\begin{equation}
	\begin{split}
		I(k,l,\nu) =  R(\nu) \times   & \left[ \Asz(k,l)~\fsz(\nu)   \right. \\
		 & + \Adust(k,l)~\fdust(\nu)  \\
		 & +  \left. \Acmb(k,l) \right], 
		 \label{equ:freq_maps}
	\end{split}
\end{equation}

where:\\
\begin{enumerate}[leftmargin=.75cm,noitemsep]
\item the Galactic dust spectrum is idealized as a uniform modified blackbody with shape $\fdust(\nu)  \propto \nu^{\beta} B_{\nu}(\nu,T)$ ;\\
\item the tSZ distortion of the CMB spectrum is modeled following the Kompaneets non-relativistic approximation: 
\begin{equation}
\fsz(\nu)  = \left[\left(\frac{h\nu}{kT}\right)\frac{exp(h\nu/{kT} + 1)}{exp(h\nu/{kT} - 1)} - 4 \right];
\end{equation}
\item each SED is  corrected  for the HFI spectral response $R(\nu)$.
\end{enumerate}

Note that the correction of each SED for the HFI spectral response includes a unit conversion factor between the HFI 100-353 GHz channels and the 545 and 857 GHz channels, that are calibrated in units of CMB temperature and intensities of a power-law SED, respectively \citep[see details in][]{Planck_2013_VIII_hfimaps,Planck_2013_IX_hfiresp}. A color correction is further applied to adapt the Galactic dust SED to the power law used to calibrate the high-energy channels. These corrections are calculated using the Unit conversion and Color Correction package that is provided with the current \planck~data release.

To perform a local regression of each component amplitude, we minimize a kernel-Weighted Least-Square (WLS) distance separating the total SED of all components, $I (k, l, n)$, from the energy distribution registered in the vicinity of each pixel of the HFI frequency maps, $y(k, l)$. Derived from the weighted likelihood formalism (see the appendix), the minimized quantity $w\chi^2(k,l)$ is weighted by the variance of the frequency maps, $\sigma(k,l,\nu)$, and spatially smoothed by a positive kernel with norm 1, $w(i,j)$:

\begin{equation}
	w\chi^2(k,l) = \sum_{i,j,\nu} {w(i,j) \frac{\left[I(i-k,j-l,\nu) -  y(i-k,j-l,\nu)\right]^2}{ \sigma(i-k,j-l,\nu)^2} } .
	\label{equ:wchi2}
\end{equation}
The minimization of $w\chi^2(k,l)$ provides us with an estimate of the searched parameter parameter vector, as follows:
\begin{equation}
	\vect{\widehat{A_{w}}}(k,l) = \left[ \begin{array}{c} \widehat{\Acmb_{,w}}(k,l) \\ \widehat{\Adust_{,w}}(k,l) \\   \widehat{\Asz_{,w}}(k,l)  \end{array} \right] = \underset{\vect{A}}{\arg\min} \left(w\chi^2(k,l)\right).
\end{equation}
It is undertaken for each pixel of the component maps, $\vect{\widehat{A_{w}}}(k,l)$, by means of a Levenberg--Marquardt algorithm \citep{Markwardt_09}.

\subsection{Component Imaging}
\subsubsection{Image denoising}

Multiscale transforms perform a sparse representation of the 2d discontinuities present in the images, thus enhancing their signal-to-noise ratio. The efficiency of the process, however,
depends on the level of correlation of the transform basis
functions to the shape of the discontinuities. For instance,
wavelet transforms best represent isotropic details, while,
compared to wavelets, curvelets are better for revealing linear
and curved edges. Because CMB features are believed to be
nearly isotropic, we chose to analyze the CMB map via its
wavelet transform. To analyze the Galactic dust and tSZ images
that plausibly hold filaments and edges, we adopted a curvelet
transform.

The WLS minimization provides us with a mechanism for extracting the undecimated \bthreespline wavelet transform \citep[][]{Starck_07} of all component maps, $\vect{{A}}(k,l)$. For each wavelet scale $a$, we split the \bthreespline wavelet function $\psi(a)$ as the sum of its negative and positive parts, $\psi_{+}(a)$ and $\psi_{-}(a)$ (see also \fig\ref{fig:bspline_kernel}) and introduce the appropriate weighting kernels in \equ(\ref{equ:wchi2}). Minimizing $\psi_{+}\chi^2(k,l) $ and $\psi_{-}\chi^2(k,l) $ provides us with an estimate of the component maps convolved with $\psi_{+}(a)$ and $\psi_{-}(a)$, and allows us to derive their wavelet transform, following:

\begin{equation}
	\wt(k,l,a) = \frac{1}{2} \times \left[\vect{\widehat{A_{\psi_{+}}}}(k,l)  + \vect{\widehat{A_{\psi_{-}}}}(k,l) \right].
	\label{equ:wt}
\end{equation}

In the same way, the variance of the wavelet coefficients can be derived from:

\begin{equation}
	d\wt(k,l,a)= \frac{1}{2}  \times \sqrt {\vect{\widehat{ {\sigma^2_{A,\psi_{+}}}}}(k,l)^2  + \vect{\widehat{ {\sigma^2_{A,\psi_{-}}}}}(k,l)^2}.
	\label{equ:dwt}
\end{equation}

To further extract the curvelet coefficients of the component maps, we normalize wavelet coefficients to their standard deviation, $\wt' = \wt/d\wt$. Subsequently, we compute a so-called curvelet transform of the first generation \citep[][]{Starck_03a} that combines a 2D \bthreespline wavelet transform with a set of ridgelet transforms that are performed within binning blocks of the wavelet detail images. Finally, the denoising of wavelet and curvelet transforms is performed via a thresholding of the coefficients whose absolute value does not exceed a few standard deviations. In the case of the CMB wavelet transform, this standard deviation is straightforwardly defined in \equ(\ref{equ:dwt}). On the other hand for the Galactic dust and the tSZ curvelet transforms, the coefficient variance depends on the orientation and it has been preliminarily estimated and tabulated by Monte-Carlo simulations of a Gaussian white noise.  In order to work with positive component maps, lower the dynamics of the reconstructed tSZ images, and reduce the relative amplitude of the thresholding artifacts, the quantities that have been effectively denoised and mapped are

\begin{equation}	
		\vect{{A_{o}}}(k,l) =  \left[ \begin{array}{c} {\rec_{ \wt}  \bar\wt[ 1 + \Acmb(k,l)]}  \\ \rec_{ \cur}  \bar\cur[{\Adust}(k,l)] \\  \rec_{ \cur}  \bar\cur[{\ln(1+\Asz)}(k,l)] \end{array} \right],
\end{equation}

where $ \bar\wt$ and $ \bar\cur$ denote the thresholded wavelet and curvelet transforms, respectively, and $\rec_{\wt}$ and $\rec_{\cur}$ denote the relative reconstruction operators.

\subsubsection{Image restoration}

Because the denoised component maps $\vect{{A_{o}}}(k,l) $ have been
estimated from the mixing of frequency maps that were
registered at various angular resolutions, they might be altered
by aliasing artifacts. In order to take advantage of the angular
resolution available at each frequency, we iteratively deconvolve
all component maps from the HFI beams by means of a
Van Cittert algorithm \citep{VanCittert_1931}. More precisely, at each iteration $i$ we associate a set of denoised frequency maps with the component maps $\vect{A_{i}}$ following \equ(\ref{equ:freq_maps}). We convolve these frequency maps with their specific beams and restore a new set of point-spread function (PSF)-convolved component maps $P*\vect{A_{n}}$ via a new combination of WLS estimate and multi-scale denoising. Starting from a first guess $\vect{{A_{o}}}(k,l) $, the Van Cittert iteration yields: 

\begin{eqnarray}
	\vect{R_{n}(k, l)} &=& \vect{{A_{o}}}(k, l) - [P * \vect{{A_{n}}} ](k, l) \nonumber \\
	\vect{{A_{n+1}}} &=& \vect{{A_{n}}}+\alpha \vect{R_{n}}(k, l)
	\label{equ:Van_cittert}
\end{eqnarray}

As proposed in \citet{Murtagh_95}, the locus of the significant (i.e. nonzero) wavelet and curvelet coefficients of $\vect{{A_{o}}}$ defines a `multiresolution support' that regularizes the algorithm. By reconstructing the residual $\vect{R_{n}(k, l)}$ from the wavelet and curvelet coefficients located within this multiresolution support \citep[see also][]{Starck_02}, we prevent any noise amplification during the Van Cittert iteration.

\begin{figure*}[h]
   \begin{center}
  \resizebox{.95\hsize}{!}{\includegraphics{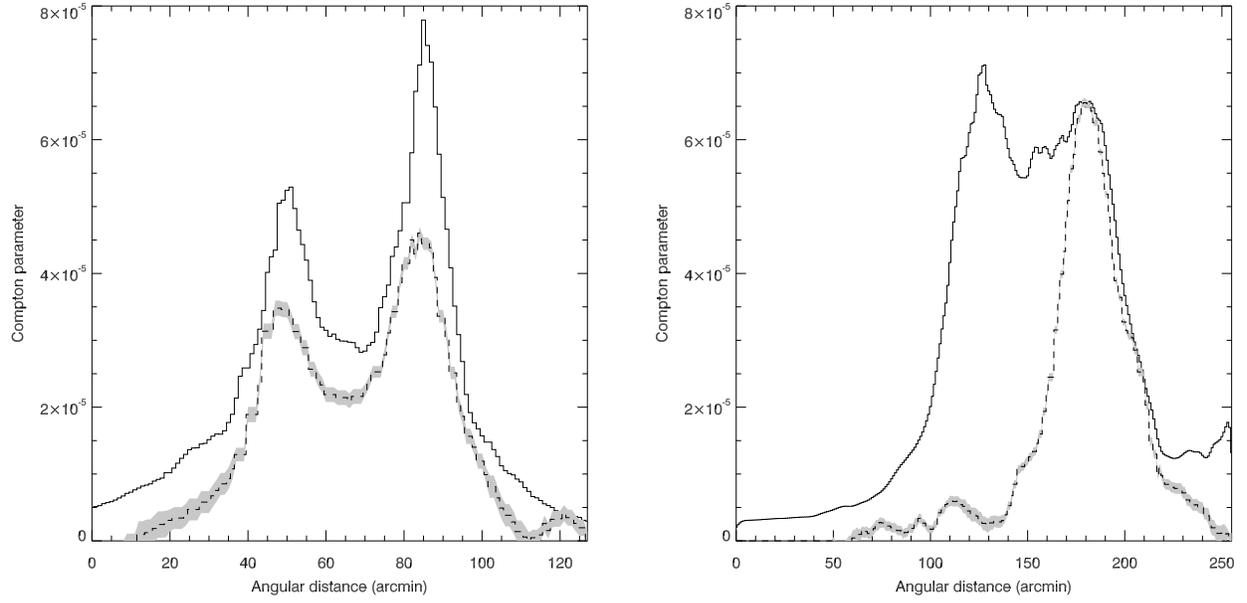}}
  \caption{Compton parameter of two SPH simulated clusters presented in this work compared with real \planck~ measurements. {{Left panel: }} $y$ parameter obtained along an image cut intercepting both maxima of the \textit{connected cluster} map displaced at $z=0.04$. The dashed line (gray area) represents a cut of the thermal SZ signal (variance), reconstructed via the MILCA algorithm toward the cluster pair A399-A401. {{Right panel: }} Same as the left panel but for the \textit{colliding cluster} displaced at $z=0.01$ and the Coma cluster.\label{Fig:a399_coma_vs_gadget}}
   \end{center}
\end{figure*}

\begin{figure*}[h]
   \begin{center}
  \resizebox{.95\hsize}{!}{\includegraphics{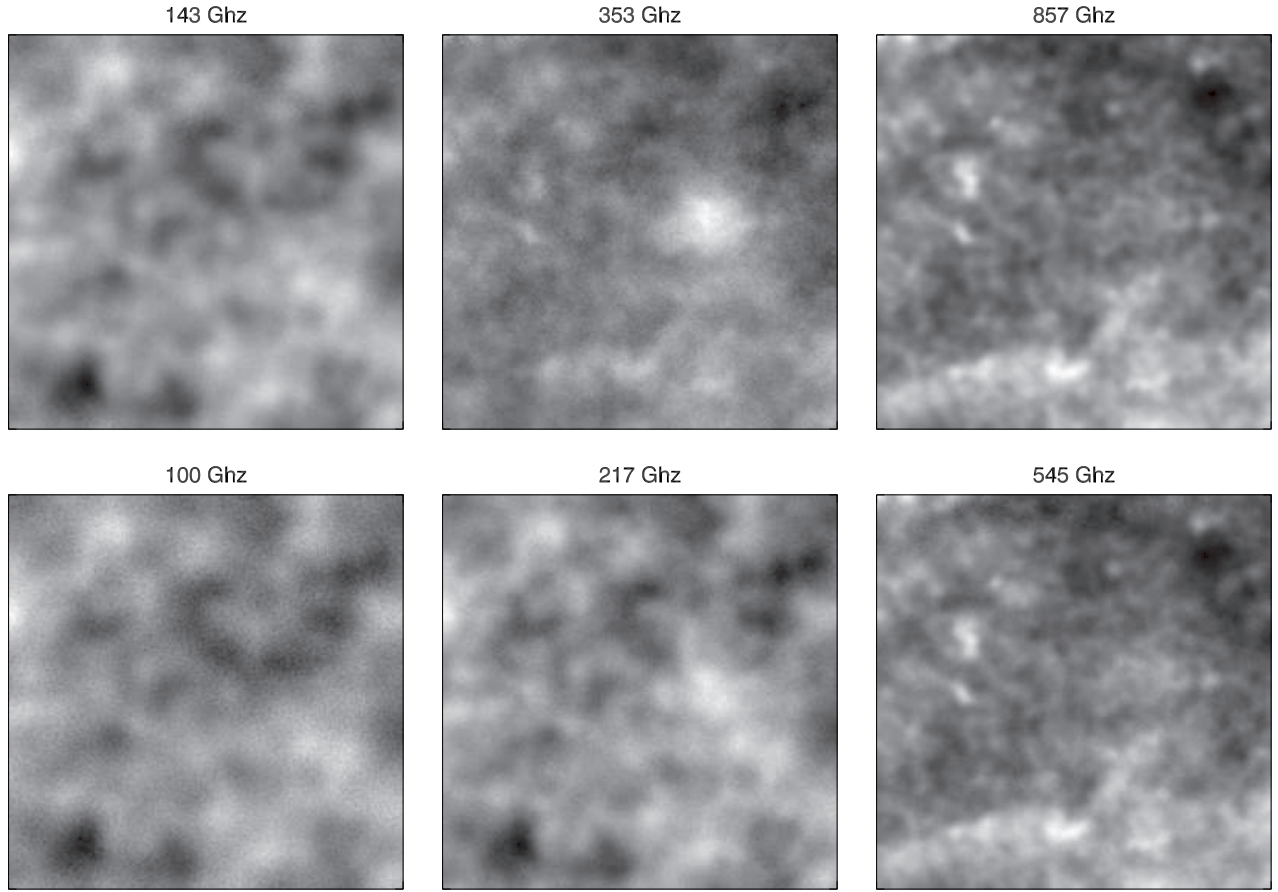}}
  \caption{Mock HFI frequency maps of the $z=0.015$ \textit{accreting cluster} positioned at $l=270^\circ$ and $b=-30^\circ$.\label{Fig:HFI_freqmaps}}
   \end{center}
\end{figure*}

\begin{figure*}[ht]
   \begin{center}
  \resizebox{\hsize}{!}{\includegraphics[]{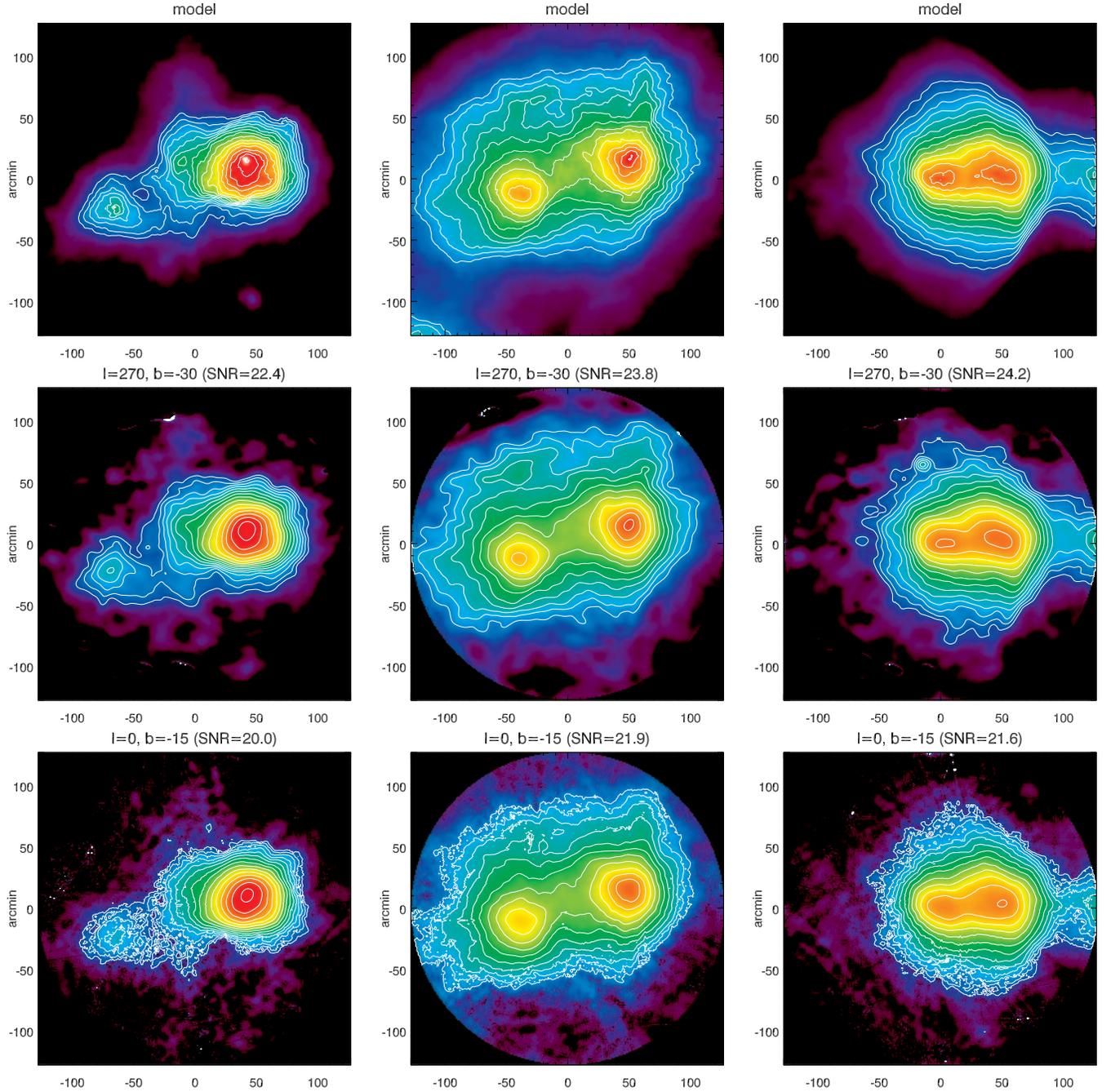}}
  \caption{{{Top panels: }} thermal SZ map of the \textit{accreting, connected, and colliding} clusters displaced at $z=0.015$, $z=0.015$ and $z=0.01$, respectively. The white isocontours are logarithmically equispaced by a factor of $2^{1/4}$. The faintest isocontour corresponds to a Compton parameter of y=$7.5 \times 10^{-6}$, except for the \textit{connected} cluster whose faintest isocontour corresponds to  $y=10^{-5}$. {{Middle and bottom panels: }} restored maps of the SZ signal for two positions in the sky characterized by a low and a high instrumental noise variance.\label{Fig:clusters_vs_snr}}
   \end{center}
\end{figure*}

\begin{figure*}[ht]
   \begin{center}
  \resizebox{\hsize}{!}{\includegraphics[]{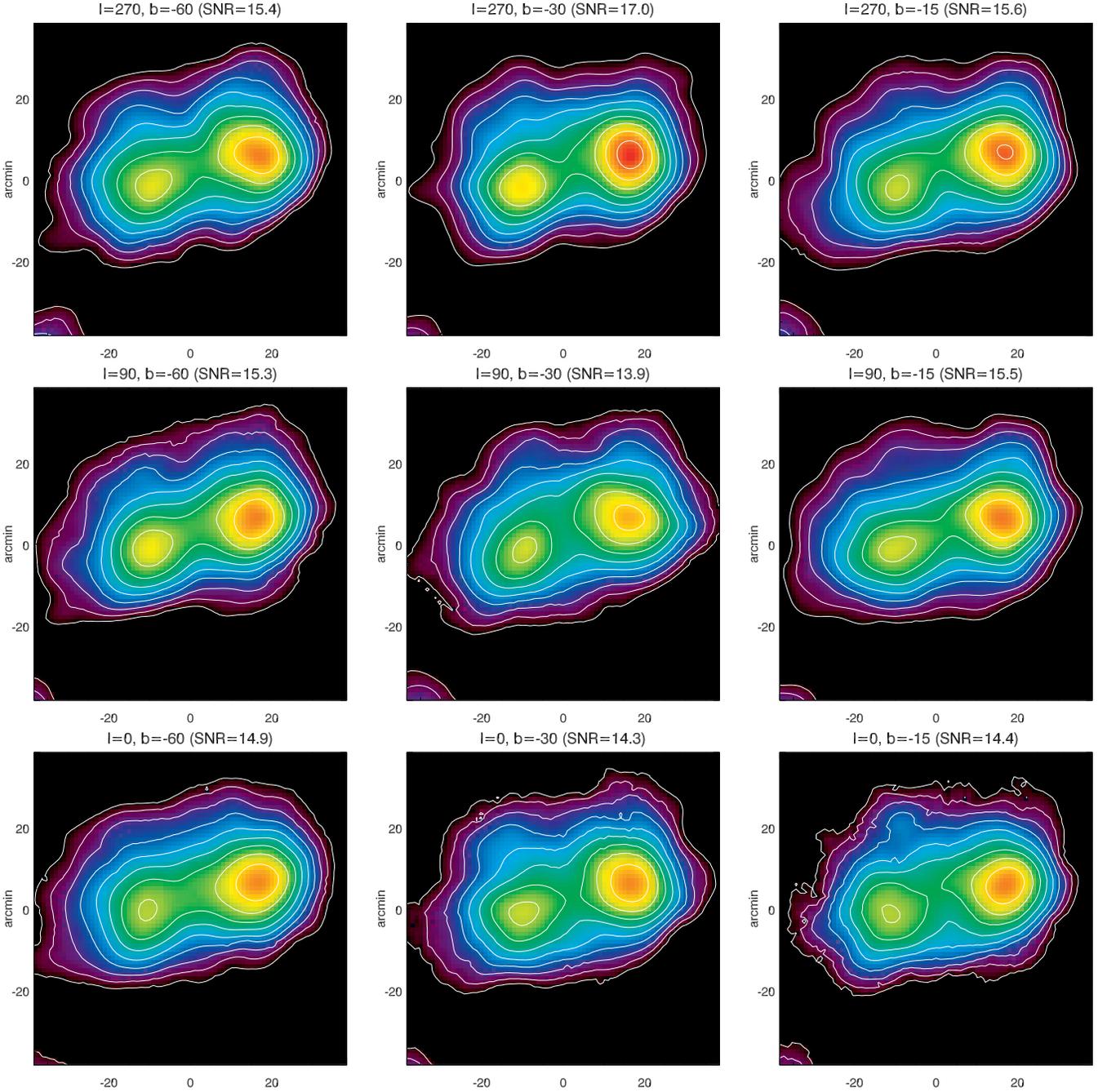}}
  \caption{Restored thermal SZ maps of the \textit{connected} cluster displaced at $z=0.05$ and positioned at various characteristic locations in the sky. The white isocontours are logarithmically equispaced by a factor of $2^{1/4}$, the faintest isocontour corresponding to a Compton parameter of $y=10^{-5}$.\label{Fig:connected_vs_position}}
   \end{center}
\end{figure*}

\begin{figure*}[ht]
   \begin{center}
  \resizebox{\hsize}{!}{\includegraphics[]{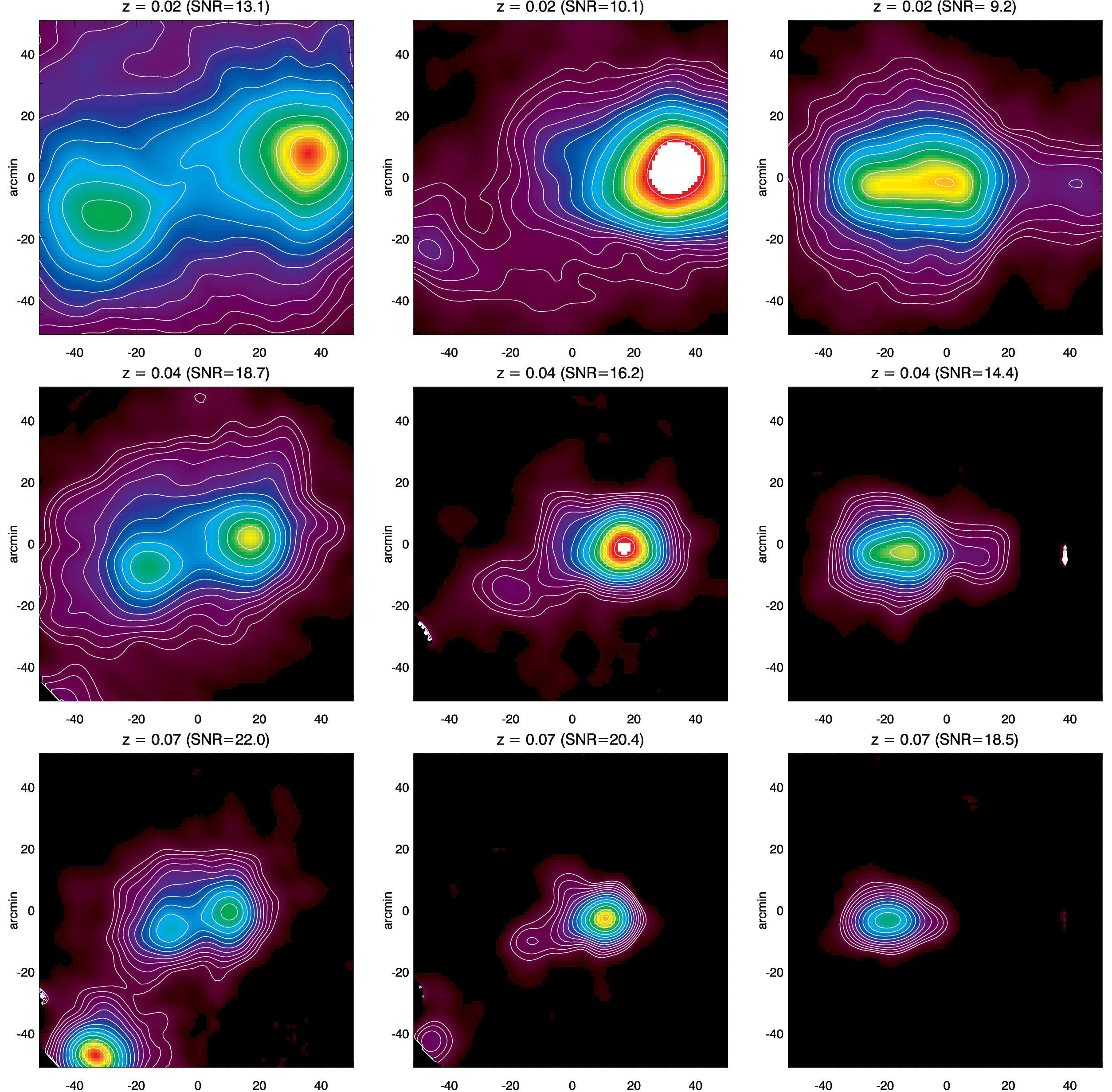}}
  \caption{{{Top panels: } Restored thermal SZ maps of the \textit{connected, accreting and colliding} clusters displaced at $z=0.02$ and positioned in the sky at ($l=270^\circ$, $b=-30^\circ$). {{Middle and bottom panels: } Same as top panels as for $z=0.04$ and $z=0.07$, respectively. White isocontours are logarithmically equispaced by a factor of $2^{1/4}$, the faintest isocontour corresponding to a Compton parameter of $y=7.5\times10^{-6}$. \label{Fig:clusters_vs_redshift}}}}
   \end{center}
\end{figure*}

\section{Mock observations of galaxy clusters\label{sect:mock_observations}}

\subsection{The thermal SZ signal as seen from SPH simulations}

In order to test the restoration algorithm detailed in the previous section, we extracted thermal SZ maps of simulated galaxy clusters. The hydro-simulations employed are carried out by the Smoothed-Particle-Hydrodynamic  code GADGET \citep{springel05}.  The details of the simulations and the physics adopted are presented in \cite{planelles.etal.14} and \cite{rasia.etal.14}. Here, we briefly list the essential points and address the
mentioned papers for a more exhaustive description. The
clusters used in the present work come from two Lagrangian
regions that were selected from a parent dark matter
cosmological box. The regions are re-simulated at a higher-mass
resolution, adding baryons either in gaseous or stellar
form. The re-simulations treat processes such as radiative
cooling; star formation and evolution; kinetic feedback by type
Ia and type II supernovae and from asymptotic Giant Branch
stars; thermal feedback from active galactic nuclei resulting
from gas accretion onto supermassive black holes \citep[see also][]{ragone.etal.13}. Masses of dark matter and gas particles are equal to $m_{\mathrm{dm}} = 8.47 \times 10^8 h^{-1} \mathrm{M}_{\odot}$ and $m_{\mathrm{gas}}=1.53 \times 10^8 h^{-1} \mathrm{M}_{\odot}$, respectively. The adopted Plummer-equivalent softening length for computing the gravitational force is set to $\epsilon = 5 h^{-1}$ kpc in comoving units up to $z=2$ then it is switched to the same value but in physical units. The minimum SPH smoothing length is $0.5 \times \epsilon$.

To prepare a few test cases that are representative of nearby clusters observed by \planck, we select three situations occuring in a rich and dense environment and displaced them at various distances from the observer. The first and second situations are two different time snapshots of the same Lagrangian region, referring to $z=0.25$ and $z=0$, respectively. The third case corresponds to a different region.

\begin{enumerate}[leftmargin=.75cm,noitemsep]
\item A \textit{massive accreting} cluster. 
With a binding mass of $M_{200}= 7 \times 10^{14} h^{-1} \mathrm{M}_{\odot}$, this cluster exhibits a rather regular morphology in the innermost regions. However, the cluster periphery is more disturbed. In particular, the object has recently accreted a satellite that is currently lying at a projected distance of 2 Mpc. 
The smaller object has a temperature of about $2$--$3$ keV, corresponding to a mass of a few $10^{14}  h^{-1}\mathrm{M}_{\odot}$. Following its cosmic evolution, we see that it will orbit on the south to the main halo with a projected impact parameter of about $500$ kpc, and will lose most of its gas content. In the cosmic time selected, the satellite and main cluster are connected by a tenuous and irregular filament with Compton parameter of $y \sim 10^{-5}$.
\item A \textit{connected} cluster pair. 
After the merger described above and after having stripped most of the gas of the subclump, the main halo is approached by another massive object of comparable mass. At redshift zero the two systems lie on the same plane of the sky and are about $1.5$ Mpc apart. By studying their evolutions, we observed that they are moving toward each other and in the future they will merge. At the current epoch, they are connected by a bright filament with a Compton parameter of $y \sim 3 \times 10^{-5}$.
\item A \textit{colliding} cluster system. 
This system is composed of multiple objects that are all located in the plane of the sky along a large-scale filamentary structure. The main cluster is at the center of the image and has a mass of $\mathrm{M}_{500}\sim 10^{15} \mathrm{M}_{\odot}$. It shows a prominent substructure at $\sim$ 50 arcmin. This is the residual of a merger with two smaller groups. The violent merger with the main halo had an impact parameter of almost zero and the gas of one of the smallest systems has been completely stripped away.   The collision direction is witnessed by a prominent bow shock that is visible as a sharp SZ edge immediately to the west of the merged objects. A third cluster, connected by a filament, is located ahead of the shock outside the field of view. The snapshot shown corresponds to $z \sim 0$, thus we cannot clearly see the future of this object. However, by looking at its past motion we presume that it will most likely produce a second major merger.

\end{enumerate}

\begin{table}
\caption{Galactic Dust and Instrumental Noise Properties Assumed at 350 GHz in the Mock HFI Frequency Maps. \label{tab:bck_par}}
\begin{center}
\begin{tabular}{cccccc}
\tableline\tableline
 \multicolumn{2}{c}{Galactic}    &  Noise      & Median    &   Dust       & Median   \\
 \multicolumn{2}{c}{coordinates} &  Standard   & Dust      &   Standard   & Dust to SZ   \\
   $l$          &  $b$           &  Deviation  & Intensity &   Deviation & Intensity  \\
   (deg)        &  (deg)         &  ($\mu$K)   & ($\mu$K)  &   ($\mu$K)  & Ratio \tablenotemark{a}  \\
\tableline
   0& -15&   99.3&  923.6&  254.6&  8.6 \\
   0& -30&   88.5&  591.8&  147.9&  5.8 \\
  90& -60&   78.8&  397.0&   74.6&  3.5 \\
   0& -60&   74.9&  104.8&   42.4&  1.0 \\
  90& -30&   72.8&  612.1&  502.6&  6.7 \\
  90& -15&   66.1& 1560.2&  422.3& 13.2 \\
 270& -60&   64.3&  257.7&   70.2&  2.5 \\
 270& -15&   60.8& 2229.5&  566.3& 22.4 \\
 270& -30&   28.2&  550.5&  135.4&  4.7 \\
\tableline
\tableline
\end{tabular}
\tablecomments{Pixel areas are 1 square arcminute. Sky regions have been sorted by decreasing noise variances.}
\tablenotetext{a}{This ratio has been measured in a region of the field of view where the Compton parameter exceeds $10^{-5}$. }
\end{center}
\end{table}

To show that these test cases may be representative of real cluster configurations, in Fig. \ref{Fig:a399_coma_vs_gadget} we compare the $y$ parameter of two of the simulations with actual \planck~measurements derived from a Modified Inter Linear Combination Algorithm \citep[MILCA; ][]{Hurier_13,Planck_2015_szmap}. The SZ map of each cluster is also visible on the top panel of Fig. \ref{Fig:clusters_vs_snr}. Fig. \ref{Fig:a399_coma_vs_gadget} demonstrates, in particular, that displacing the \textit{connected cluster} at $z=0.04$ yields an angular separation between the two cluster peaks that is comparable to the separation of both components of the cluster pair A399-A401. Moreover, displacing the  \textit{colliding cluster} at $z=0.01$ leads to a Compton parameter decrement that matches its true analog that is measured across the most prominent shock front in Coma \citep[see also][]{Planck_intermediate_X_coma}.

\subsection{Mock HFI frequency maps}

To emulate HFI observations of our simulated clusters, we choose a few characteristic regions of the sky in terms of thermal SZ signal-to-noise ratio and predict the 2D signal to be registered in each HFI channel by adding our tSZ maps to the expected specific intensities of the CMB and Galactic dust components. For each HFI channel and sky region, frequency maps $I(k, l, \nu)$ are convolved with the HFI beam and added to a noise realization that matches the noise variance and power spectrum registered to the cluster scales in the raw data. The amplitude of the CMB SED is extracted from the Planck CMB map that is
reconstructed using the Spectral Matching Independent Component Analysis (SMICA) component separation algorithm \citep{Planck_2013_XII_components}, while the amplitude of the Galactic dust SED accommodates the dust optical depth mapped at 353 GHz as a result of the \planck~ all-sky model of thermal dust  \citep{Planck_2013_XI_tdustmodel}. For simplicity, a constant spectral index, $\beta$ = 1.8, is assumed for the dust SED, consistent with the first all-sky modeling from \planck~ and IRAS data \citep{Planck_early_XIX_tdustmodel}. An example of mock HFI frequency maps of the \textit{colliding} cluster system is shown for ($l=270^\circ$, $b=-30^\circ$) in Fig. \ref{Fig:HFI_freqmaps}, while a few properties of the Galactic dust and instrumental noise included in the 350 GHz maps are summarized in Tab.\ref{tab:bck_par} for each simulated sky position. The dust intensity in particular mostly depends on the Galactic latitude, and is shown to dominate the positive side of the thermal SZ signal ($f_{\nu} > $ 217 GHz). 

\subsection{SZ Map restoration}

 Figures \ref{Fig:clusters_vs_snr}--\ref{Fig:clusters_vs_redshift} present the restoration of simulated SZ maps for a soft thresholding, 

\begin{equation}
\bar\wt = \text{sgn}(\wt) \left(|\wt| - \lambda\right)_{+}
\end{equation}
of the wavelet and curvelet coefficients at $\lambda = 1.5~\sigma$, after 3 Van Cittert iterations performed with $\alpha=0.25$ (see Equ. \ref{equ:Van_cittert}). For each map, the restored signal, $\Asz$, is shown together with the signal-to-noise ratio:
 
 \begin{equation}
SNR =10 \log \frac {\| \Asz \|}{\| \Asz - \widehat{\Asz}\|}.
\end{equation}

{\textit{Effect of the instrumental noise variance.\\ }}
Fig. 4 shows the tSZ maps of our test clusters positioned at low redshift, in a couple of sky regions corresponding to the lowest and highest variance of the instrumental noise in Tab. 1. In this figure, all characteristic features with a typical Compton parameter of $y \ge10^{-5}$ are bright enough to be restored. More specifically, both components of the \textit{accreting} cluster are recovered for each sky region, though the SNR decrement makes it difficult to recover the true shape of the tenuous filament connecting the two clusters at ($l=0^{\circ}$, $b=-15^{\circ}$). The extended bow shock escaping from the \textit{colliding} clusters is also detectable for each sky region, together with the filament connecting these clusters to their third companion. \\

{\textit{Effect of the instrumental noise variance and dust intensity.\\ }}
In Fig. 5, the tSZ signal of the \textit{connected cluster} displaced at $z=0.04$ is mapped for the sky positions detailed in Tab. 1. The cluster system is restored with an SNR larger than 14 for each sky region. In particular, the position of the cluster peaks and the orientation of the connecting filament are preserved in any case. The image SNR turns out to be roughly correlated with the instrumental noise variance regardless of the Galactic latitude, which suggests that it essentially depends on statistics. \\

{\textit{Effect of the angular distance\\ }}
A few tSZ maps of our test clusters that were displaced at higher redshifts are finally shown on Fig. 6. The redshift values of 0.02, 0.04 and 0.07 correspond to two successive decrements of the angular distances by a factor of two. Separated by two PSF radii, both components of the \textit{connected} and \textit{accreting} clusters are well-resolved even at these redshifts. The shock front is also clearly visible at $z=0.02$, while its companion filament is detected up to $z=0.04$.

\begin{figure}[t]
   \begin{center}
  \resizebox{.95\hsize}{!}{\includegraphics{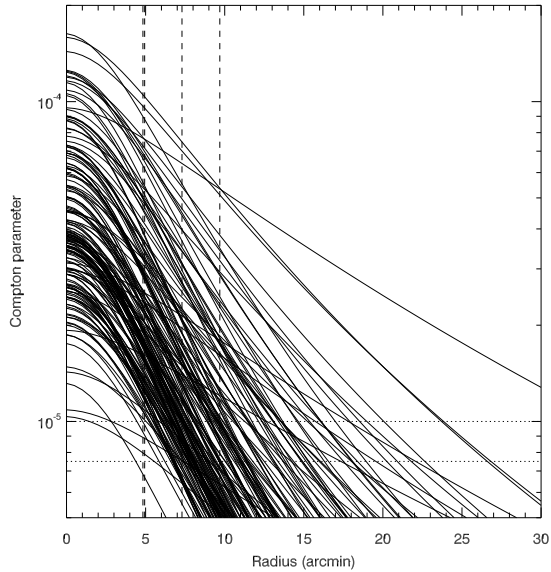}}
  \caption{Radially averaged Compton parameter of a nearby sample of galaxy clusters detected with a threshold higher than $S/N=5$ in the  \planck~ catalog.  Apparent Clusters radii, $r_{500}$, have been selected to exceed the PSF width for each HFI frequency map. This modeling assumes the radial pressure structure proposed by \citet{Arnaud_10}, and a 2D convolution by a mixture of the Gaussian Planck beams weighted by the expected SZ fluxes at each frequency. {Horizontal lines}: faintest isocontour levels in Fig. \ref{Fig:clusters_vs_snr} to \ref{Fig:clusters_vs_redshift}. {Vertical lines}: PSF widths expected in the four HFI frequency maps characterized by a significant thermal SZ signal. 
 \label{Fig:yszcat_profiles}}
   \end{center}
\end{figure}

\subsection{Perspectives on real observations}

In the previous section we showed that ICM anisotropies are accurately restored by our algorithm, provided that they are spatially resolved in each HFI frequency map and located in cluster regions where the Compton parameter, $y$, exceeds $10^{-5}$. As shown in Fig. \ref{Fig:clusters_vs_snr}, the situation is more critical for $y$ values that are lower than $10^{-5}$, since the restoration depends on statistics and dust emissivity for a given region of the sky. In practice, a severe detection threshold (3$\sigma$ or more) would allow us to isolate ICM features down to $y$ values of  $5\times10^{-6}$ in most of the sky regions, in particular when they are extended enough to be analyzed and averaged out on several PSF sizes.

Though not intended to be considered as absolute limits for each specific target, these values provide us with a coarse assessment of the potential of the presented algorithm in the 2015 release of \planck~ data. To predict a number of appropriate targets,  we selected a sample of galaxy clusters in the \planck~ catalog of SZ sources whose $r_{500}$ radii exceed the PSF size of each HFI frequency map, and whose detection thresholds, S/N, are higher than 5 in the sense of the Multifrequency Matched Filter detection algorithm detailed in \citet{Melin_06}. Fig. \ref{Fig:yszcat_profiles} exhibits the radial profiles of the $y$ parameter expected toward these clusters if one assumes a pressure structure that follows the average profile measured for a representative sample of the X-ray cluster population \citep{Arnaud_10}. The cluster radii, $r_{500}$, and integrated-pressure, $\mathrm{Y}_{5r_{500}}$, follow the scaling relation proposed by the Planck collaboration to constrain cluster masses from SZ measurements \citep{Planck_2013_XX_cosmo}. The radially averaged Compton parameter exceeds $10^{-5}$ for 57 clusters at a distance from the center that exceeds the PSF size of each HFI frequency map. Lowering our selection criterion, at variance, to a $y$ parameter of $7.5\times10^{-6}$ would yield 81 clusters for which ICM anisotropies would be detectable beyond the PSF size of each HFI frequency map. 

Beyond the brightest central regions of clusters, matter filaments also connect the component of cluster systems, and might become detectable at a larger radii than expected within isolated, spherically symmetric clusters. A typical sample of targets for such studies is composed of 18 pairs of clusters from the \planck~ catalog, detected with an S/N threshold that is higher 5, showing an angular separation that is lower than $1^{\circ}$ and a redshift interval that is lower than 0.01.

\section*{Conclusion}

	We presented a multi-scale algorithm aimed at restoring the thermal SZ maps of extended galaxy clusters observed with the \planck~ HFI bolometer arrays. The demixing of the thermal SZ
signal with respect to the CMB and thermal dust is tackled
through locally weighted likelihood maximizations, which can
be seen as a generalization of kernel smoothing to the
multivariate case. The restoration of the component maps is performed via a Van Cittert deconvolution that is restricted to the most significant wavelet and curvelet coefficients of these components. 
	
	The algorithm performance has been tested against mock observations of galaxy clusters positioned at various characteristic positions in the sky. Though unavoidably idealized\footnote{Typical real restoration of the SZ signal might also include a modelling of the expected spatial variation of the dust spectral index, and a masking of known point-like sources that hold a specific SED.}, these observations mimic the noise variance and power spectrum of the full \planck~ mission, and also hold CMB or dust SED and anisotropies that are representative of the extended sky emissions at high Galactic latitudes \citep[$|b| > 20$; see, e.g.][]{Boulanger_96,Planck_2013_XI_tdustmodel}. They show us, in particular, that \planck~ should allow us to detect filaments in the cluster peripheries and large-scale shocks in colliding galaxy clusters that feature a favorable geometry. The \planck~ catalog of SZ sources includes about 60 bright and well-resolved galaxy clusters for which such features might be detected. It also holds a number of spatially resolved cluster systems, which are appealing candidates for searching for connecting filaments. The unique algorithm input being a set of radio frequency maps with their characteristic variances and beams, other space or ground based observations might be added to the HFI data in order to further improve the angular resolution and SNR of the restored SZ maps. 
	
\acknowledgments

We wish to thank Giancarlo De Gasperis, Guillaume Hurier, and Juan Macias-Perez for fruitful discussion about the \planck~data analysis. We also thank the referee for her/his constructive comments which helped us to improve our manuscript. We acknowledge the use of image analysis routines of the Interactive Sparse astronomical data Analysis Package (ISAP) that was developed in the CosmoStat laboratory at CEA Saclay. We are greatly indebted to the whole Dianoga team, who produced the simulations used in this work (PIs. Stefano Borgani, Giuseppe Murante, Klaus Dolag). The mock observations presented in this work mimick real observations that were obtained with \planck, an ESA science mission with instruments and contributions that are directly funded by ESA Member States, NASA, and Canada. H.B. thanks the University of Michigan, where this work was initiated, for its hospitality. P.M. acknowledges support by grant NASA NNX14AC22G. E.R. acknowledges support by FP7-PEOPLE-2013-IIF (Grant Agreement PIIF-GA-2013-627474) and NSF AST-1210973.

\appendix

\section{Spatially weighted likelihood and $\chi^2$ estimates.}

Let us assume that a spatially variable data set, $x_t$, sampled along coordinate $t$, is the realization of the probability density function $p(x_t | \thetavec(t))$. An asymptotically unbiased and efficient estimate of the underlying parameter vector $\thetavec(t_o)$ might be provided by the maximization of its log-likelihood function within some coherence region $\reg$, perhaps adjacent to $t_o$:

\begin{equation}
\widehat{\thetavec_{L}}(t_o) = \underset{\thetavec}{\arg\max} \sum_{t \in \reg} log\left[p(x_t | \thetavec)\right]
\label{equ:L}
\end{equation}

This estimate, however, relies on the assumption that $\thetavec(t_o)$ is locally stationary within $\reg$, whose morphology and extension may be unknown a priori. Adopting a Bayesian point-of-view, the weighted log-likelihood approach \citep[e.g.][]{Hastie_86, Fan_98} introduces a spatially variable penalization $w(t)$ that aims to lower any relative entropy loss associated with spatial variations of $\thetavec(t)$, and thus reduce the risk of the maximum likelihood estimate \citep{Wang_06}:

 \begin{equation}
 	\widehat{\thetavec_{wL}}(t_o) =  \underset{\thetavec}{\arg\max} \sum_{t \in \reg} w(t)~log\left[p(x_t | \thetavec)\right] \label{equ:weighted_L}
 \end{equation}

Weighted likelihood estimates have already been proposed to denoise images that were altered by various kinds of parametric noise models, yielding specific local or non-local smoothing kernels \citep{Polzehl_06,Deledalle_09}. They typically reduce the variance of the likelihood estimates, at the potential cost of introducing a bias related to the spatial variations of $\thetavec(t)$ relative to $w(t)$ \citep[][]{Fan_98,Eguchi_98}. In the present work, this scheme is applied in order to spatially smooth the local estimates of three extended component maps, $\thetavec(t)$, which combine linearly with each other to provide us with \planck~HFI frequency maps. To lower the weighted likelihood bias, a key issue is to choose appropriate weights that tend to gather regions where the underlying parameters $\thetavec(t)$ are likely to be uniform. It is the thresholding of wavelet and curvelet coefficients that are derived from convolutions of $\widehat{\thetavec_{wL}(t_o)}$ with a the positive and negative part of wavelet function, that allow us to achieve this goal a posteriori. 

To adapt the weighted likelihood estimate to the \planck~HFI measurement of slowly variable flux, $\mu(\thetavec(t))$, we may assume an additive gaussian noise with variance $\sigma_t$. Introducing  $p(x_t | \thetavec)=N(x_t-\mu(\thetavec), \sigma_t)$ in Equ. (\ref{equ:weighted_L}) leads to a weighted least squares minimization, as follows:
  
 \begin{equation}
 	\widehat{\thetavec_{w\chi^2}}(t_o) =  \underset{\thetavec}{\arg\min} \sum_{t \in \reg} w(t) \left[ {\frac{[x_t-\mu(\thetavec)]^2}{{\sigma_t^2}}} \right] \label{equ:HFI_wls} 
 \end{equation}

\hspace{1cm}

In this way and provided that $\mu(\thetavec)$ is monotonous, deriving the above summation with respect to $\thetavec$ yields $\mu (\hat\thetavec) = \frac{\sum{w(t)  / {\sigma_t}^2} x_t}{\sum{{w(t)/ {\sigma_t}^2}}}$. $\mu (\hat\thetavec)$ and $\hat\thetavec$ are thus reached via a smoothing of the data set $x_t$ with kernel $w(t)$, revealing us in turn any local linear perturbations of $\mu (\thetavec(t)) - \mu (\hat\thetavec(t_o))$ as a function of the spatial variations of the underlying component maps. Moreover, introducing the change of variables $\sigma_t^{\prime} = \sqrt{w(t)} \sigma_t$, Equ. (\ref{equ:HFI_wls}) is formally equivalent to an unweighted $\chi^2$ estimate of $\thetavec$, now assuming that $p(x_t | \thetavec)=N(x_t-\mu(\thetavec), \sigma_t^{\prime})$, which allows us to derive $\widehat{\thetavec_{w\chi^2}(t_o)}$ and its variance from any $\chi^2$ minimization algorithm. In our work, this minimization is extended to the six \planck~HFI frequency maps, yielding Equ. (\ref{equ:wchi2}). Given the specific models and smoothing kernels applied, a Levenberg--Marquardt algorithm ought to be more robust than a linear regression, especially considering the possibility of adding physical priors such as the positivity of two of the searched parameters. 
       	
\bibliography{planck_gadget_apj}

\end{document}